\documentclass[aps,prbbib,floatfix,floats,twocolumn]{revtex4}
\usepackage{graphicx,latexsym}
\usepackage{epsfig}
\usepackage{amsmath}
\usepackage{color}
\usepackage{bm}

\begin{document}

\title{Thermal memory fading by heating to a {\it lower} temperature: experimental data on polycrystalline NiFeGa ribbons and 2D statistical model predictions}

\author{F. \c Tolea, M. \c Tolea*, M. V\u aleanu}

\address{
 National Institute of Materials Physics,
POB MG-7, 77125
 Bucharest-Magurele, Romania.}

\begin{abstract}

Shape memory alloys are known to memorise one -or several- temperatures at which the martensite-austenite transformation was stopped before completion in the past, the memory manifesting as specific dips in subsequent calorimetric scans. Previous studies have shown that this memory can be erased by heating to {\it higher} temperatures than the ones previously recorded. In this paper, we study a distinct memory fading effect which takes place by heating to a {\it lower} temperature. This effect is reported in NiFeGa as polycrystalline ribbons, the alloy being initially studied as bulk for which the thermal memory effect was not found. If, after an initial incomplete heating up to $T_1$ one performs a second incomplete heating up to $T_2<T_1$, a new calorimetric dip appears at $T_2$, as expected, while less expected was that the dip corresponding to $T_1$ reduces in amplitude or even vanishes (if the arrest at $T_2$ is repeated). The memory fading effect is more clear for small differences $T_1-T_2$ and less obvious or absent for large ones. The second part of the paper employs a statistical 2D model, which associates the memorized temperatures with a depletion of certain martensite plates sizes, and also supports the memory fading effect.

\vskip 0.5cm
\begin{it}
 Random filling of a surface with squares
\end{it}

\end{abstract}

\maketitle

\section{Introduction}

The shape recovery of the -there after named- shape memory alloys (SMA) is well understood to rely on the small enthalpy difference between the austenite and martensite, the non-diffusive nature of the phase transition, and the versatile geometrical accommodation of the martensite variants. As such, the material  may undergo the reversible phase transition rather then accumulate elastic stress or create defects - when acted upon.

However, the SMA proved that they can remember not only shapes, but also temperatures at which the austenite-martensite phase transition was stopped before completion in the past (see, e.g. \cite{Airoldi1,Mad-Scripta2,R-AACTA,R-AAIM,JALCOM1,Cui,Liu}). Unlike the shape memory effect, the thermal memory effect (TME) in SMA is less understood, proposed scenarios including the redistribution of accumulated stress \cite{Mad-Scripta2,R-AACTA,R-AAIM}, or geometrical constrictions effects \cite{Tolea1}.
Within a brief definition, TME means that if the martensite-austenite transformation is stopped before completion (or "arrested", at a temperature $T_A$), then after the sample is cooled back into the martensite state, if one performs one final, complete martensite-austenite transformation,  a specific dip will appear in the calorimetric signal at a temperature close to $T_A$, proving that the alloy has temperature memory.

The TME is not restricted to SMA, being also found in FeRh which has a first order magnetic transition \cite{Kakeshita1}, suggesting that it may rely on more general principles than the particularities of the martensitic transformation. Apart from the fundamental relevance towards a better understanding, the effect may find also important practical applications \cite{Cui,Tang}.

Further studies on the TME (see, e.g. \cite{Airoldi1,WANG_SM,R-AACTA2,Wang-IJSNM,Zhou}) have systematically reported on:
\begin{itemize}
  \item[(i)] Memorising of a single, but also of multiple arrest points if the arrest temperatures are (chronologically) in decreasing order $T_1>T_2>....>T_n$.
  \item[(ii)] Erasing the memorized arrest point(s) after heating to a temperature superior to the memorized one(s). If multiple temperatures are memorized, as described above at (i), then after a heating to, say, $T_E$, all the dips corresponding to $T_i<T_E$ are erased, while those corresponding to $T_i\geq T_E$ remain memorized.
  \item[(iii)] Increase of the effect by repetition - i.e. the "hammer effect"- meaning a more clear dip in the calorimetric scan if the arrest procedure is repeated for the same temperature.
\end{itemize}

In the present paper, we investigate a different effect, which is a memory "fading" taking place by heating to a {\it lower} temperature [distinct from the property (ii) described above, according to which memory is erased by heating to a {\it higher} temperature].
In support of this effect we bring experimental data on NiFeGa polycrystalline ribbons, and also statistical arguments using a phenomenological 2D model.

It is important to mention that in our opinion some previous experimental data reported in literature (see, e.g. Fig.7 form \cite{Wang-IJSNM}, on TiNiCu thin films or Fig.2 form \cite{Zhou} on MiMnGa based bulk alloys) may be interpreted as evidence of memory fading by heating to lower temperatures, but have not been given this particular interpretation.

\section{Experimental results for $NiFeGa$ polycrystalline ribbons}

\begin{figure}[ht]
\centering
\vskip 0.1cm
\hskip 0.1cm
\includegraphics[scale=0.9]{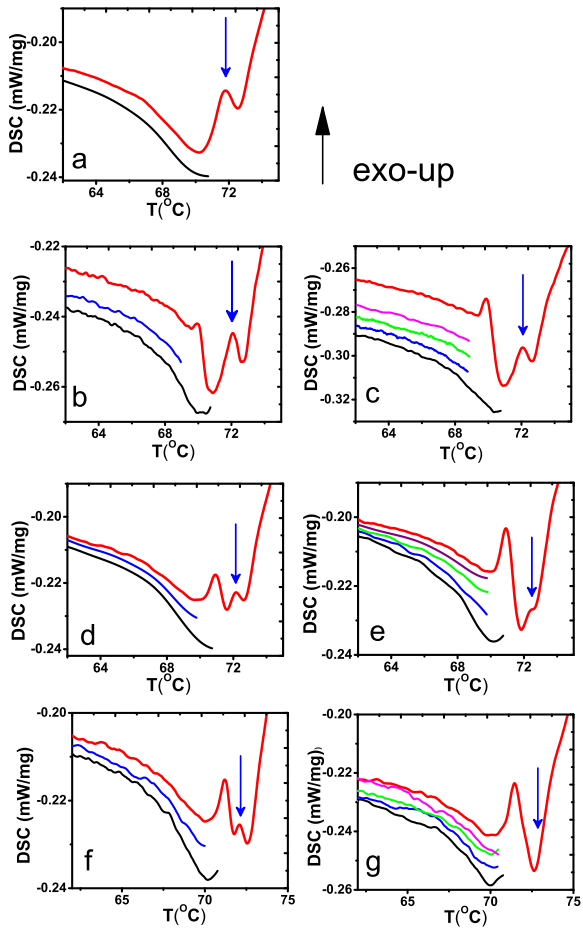}
\vskip -10.5cm
\caption{{\it Complete} DSC scans (red curves) corresponding to the martensite-austenite phase transition in NiFeGa ribbons, recorded after one or several {\it incomplete} -i.e. arrested- previous martensite-austenite phase transitions (curves with other colors then red): (a) one arrest at $T_1=71^oC$, (b) one arrest at $T_1$ and one at $T_2=69^oC$, (c) one arrest at $T_1$ and three at $T_2$, (d) one arrest at $T_1$ and one at $T_3=70^oC$, (e) one arrest at $T_1$ and three at $T_3$, (f) one arrest at $T_1$ and one at $T_4=70.5^oC$, (g) one arrest at $T_1$ and three at $T_4$. The blue arrows indicate the position of the dips associated with $T_1$.}
\end{figure}

In this Section, we present the results of calorimetric measurements (DSC) corresponding to the martensite-austenite phase transition in NiFeGa polycrystalline ribbons. This alloy was previously studied as a cheaper and less brittle alternative to  NiMnGa \cite{Chernenko} - while keeping a good level of magnetic properties as a ferromagnetic shape memory alloy (FSMA) \cite{Oikava}.
In previous works, NiFeGa exhibited the possibility of transformation temperatures or enthalpies tuning \cite{Tolea_JALCOM}, showing a rich magnetoresistive behavior \cite{IEEE}. Our focus here is to characterize the thermal memory properties, i.e. the way in which this alloy memorizes temperatures at which previous transformations have been stopped before completion in the past.

First, we searched for memory properties of the bulk NiFeGa motivated by the recent discovery of the TME effect in MiMnGa \cite{Zhou}. However, our bulk samples did not show the specific dips associated with temperatures memorizing after incomplete reverse phase transition (not shown). Not finding the effect in bulk, we further sought it in polycrystalline ribbons, prepared by rapid quenching of the melt, as described in the following.
For this purpose, we have chosen from the NiFeGa compositions studied in \cite{Tolea_JALCOM} the compound Ni56Fe16Ga28   showing narrow and well defined transformation peaks.
Ingots of the mentioned nominal composition have been prepared from high purity elements, by arc melting under argon protective atmosphere and subjected to a thermal treatment in high vacuum for 25 h at 1223 K, followed by a quenching in iced water. The ingots were inductively melted in quarts tubes under argon atmosphere and rapidly quenched by melt spinning technique (Buhler Melt Spinning Device). Long ribbons of about 30 $\mu m$ thickness and 3 mm width were obtained (copper wheel velocity of 20 m/s, 50 kPa Ar overpressure, crucible nozzle diameter of 0.5 mm). The as-quenched (AQ) ribbons were subjected to thermal annealing for 60 minutes at 673 K.
DSC scans have been performed at 10K/min scanning rate under He protective atmosphere (using a Netzsch DSC 204 F1).

Fig.1a shows that the polycrystalline NiFeGa prepared as ribbons exhibits thermal memory, unlike the bulk counterpart.
A specific dip appears in the calorimetric scan -indicated by the blue arrow- after a previous incomplete reverse transformation was arrested at $T_1=71^oC$  (black curve). Next, we investigate the proposed scenario of memory fading, by looking at what happens if, subsequently, another lower temperature is recorded. Fig.1b shows that consecutive thermal arrests performed at $T_1$ and then at $T_2=69^oC$ lead to clear memorizing of both temperatures. Repeating the arrest at $T_2$ three times accentuates the dip corresponding to this temperature (hammer effect) while a reduction in amplitude of the dip corresponding to $T_1$ is noticed.

In the following we keep the initial arrest at $T_1$ but bring the second arrest temperature progressively closer. Fig.1d shows that if the second arrest is performed one time at $T_3=70^oC$, a considerable reduction of the dip corresponding to $T_1$ can be noticed, together with the expected second dip at $T_3$.
Repeating the arrest at $T_3$ three times leads to almost complete erasure of the memory corresponding $T_1$, only a "shoulder", rather than a dip, being still prezent (Fig.1e). Finally, if the second arrest is performed at $T_4=70.5^oC$ the dip at $T_1$
drastically reduces after a single procedure and completely vanishes after three arrests at $T_4$, showing that the memory fading effect is stronger if the second arrest temperature is closer to the first one.

\section{Phenomenological 2D model and simulations results}

 \begin{figure}[ht]
\centering
\vskip 0.1cm
\hskip 0.1cm
\includegraphics[scale=0.45]{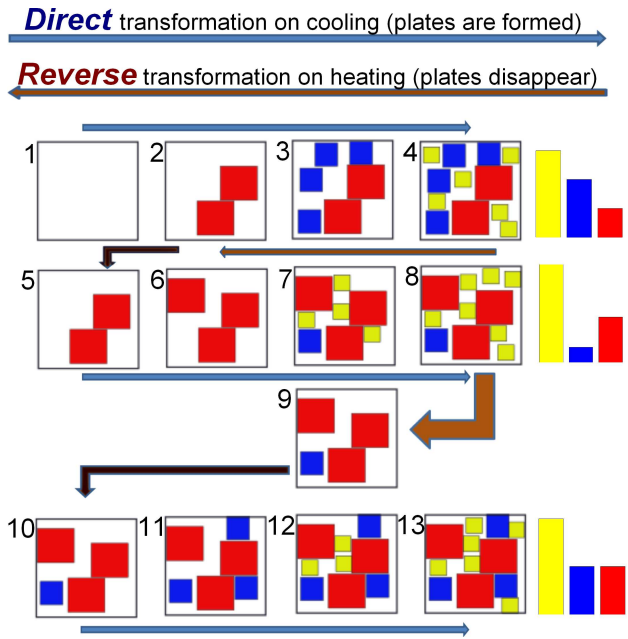}
\vskip -0cm
\caption{Schematic representation. Plates are assumed to form -direct transformation- in decreasing sizes order and to disappear -reverse transformation- in increasing sizes order.
 {\it First row, left to right (Panels 1$\rightarrow$4)}, a "normal" direct transformation is depicted, taking place by formation of progressively smaller plates (red, blue and yellow), as the temperature is lowered $T_1>T_2>T_3>T_4$. {\it First row, right to left (Panels 4$\rightarrow$2)} an incomplete reverse transformation. Yellow and blue plates transform back, while the red ones remain untransformed. {\it Second row, left to right (Panels 5$\rightarrow$8)} in the subsequent direct transformation new red plates may form ({\it Panel 6}). This "anomalously" high number of big plates makes it difficult for the intermediate blue ones to geometrically fit, so most of them only grow to a yellow size ({\it Panel 7}). {\it (Panels 8$\rightarrow$9)} let us now assume that the next reverse transformation only transforms back the yellow plates. {\it Lower row, left to right (Panels 10$\rightarrow$13)} the untransformed blue plates plus the (statistically possible to form) new ones lead to an overall greater number of blue plates. If the depletion of blue plates in Panel 8 is a memory effect, their increased number in Panel 13 is a memory "fading". Bar charts indicate the plates sizes distributions in the Panel nearest to them.}
\end{figure}

In the following, we give the predictions of a 2D phenomenological model, first proposed in \cite{Tolea1}.
The model assumes that the austenite-martensite phase transition takes place by successive formation of finite size square-shaped plates, which nucleate randomly in the untransformed area. At the beginning of the phase transition -at high temperatures- bigger plates are formed, while in the final stages -at low temperatures- smaller plates are formed. Such a decrease of plates sizes with temperature is supported by experimental evidence \cite{Jost}, and can be given also thermodynamic arguments, as discussed below. This scenario is depicted in Fig.2, first row from left to right, where a "normal" austenite martensite phase transition is schematically described, with the distribution of plates sizes indicated by the bar chart on the right.

The experimental equivalent of our plates would be a fully grown self-accommodation unit. Its internal structure (as for instance the twining) is neglected at this point, while it would be insightful  to consider it in further developments of the model - in particular this will also mean to account for more general shapes, such as rectangular, lameltransformationlar, etc.

There is also a key assumption made for the reverse transformation, namely that the smaller plates transform back first due to the largest surface to volume ratio making them the first to become thermodynamically unstable. If we look again at Fig.2, first row, this time from right to left, we have a reverse transformation in which the yellow plates disappear first (i.e. transform back to austenite), and then the blue ones.
Let us now assume that this reverse transformation is stopped before completion -as happens in the thermal memory experiments- leaving the red plates untransformed. Then, the subsequent direct transformation will start form this pre-defined puzzle of big size plates (Fig.2, second row). More big size (red) plates now statistically have a chance to form but consequently the intermediate size plates have more statistical chances to nucleate in places where they cannot grow to the "blue" size due to geometrical constrictions, and they only grow to a "yellow" size (Panel 7). Panel 8 in Fig.2 and the nearby bar chart show such a possible situation, with a pronounced depletion of intermediate size blue squares. This depletion of blue squares can be interpreted as a thermal memory, since it directly results from stopping the previous reverse transformation at the temperature at which the blue plates transformed back.

Next, we propose a memory "fading" scenario, which within our model should translate as a reduction of the depletion of blue plates. For this, we consider another incomplete reverse transformation, this time stopped at a lower temperature. As seen in Panel 9 of Fig.2, we now assume that just the yellow plates have been transformed back, while the blue ones remain untransformed and are part of the plates puzzle from which the next direct transformation starts (last row of Fig.2, Panel 10). Red plates are very unlikely to form, as they cannot fit. The formed nuclei have statistically two possibilities: to form in places where they can only grow to yellow sizes, or to form in places where they can grow to blue sizes. When averaging over a large number of configuration, both cases will occur, but the net result will be an increase of the blue plates number, thus a reduction of the depletion (we must keep in mind also that the newly formed blue plates add to the existing one, so their number can only grow in this scenario), which can be interpreted as a "memory fading".

These general arguments are tested by involved numerical simulations presented in Fig.3.
We shall not repeat all the numerical details -which are rather elementary and given in \cite{Tolea1}- but mention just a few.
During the real-time simulations, isothermal nucleation rate is assumed for the new formed squares (there is an interesting on-going debate regarding the isothermal or athermal nature of the martensitic transformation \cite{PRL2001,Scripta_Iso,Scripta_Iso_Lee,Scripta_Salas,Dubi} ) : $J_{A\rightarrow M}(T)=J_0 S ~~ exp ( -\frac{\Delta\Omega_c(T)}{k_BT})$, $S$ being the untransformed surface, and $\Delta\Omega_c(T)$ the critical free energy. If the enthalpy difference per unit area between martensite and austenite in the vicinity of the equilibrium transition temperature $T_0$ is in the linear approximation $\Delta F=\epsilon (T-T_0)$ and the interface energy penalty per unit length is $\sigma$, this leads to a total free energy difference $\Delta\Omega(T)=L^2\cdot \Delta F+4L\cdot\sigma$. The condition $\frac{\partial}{\partial L}\Delta\Omega_{A\rightarrow M}=0$ yields the critical square side $L_c=-2\sigma / \epsilon (T-T_0)$ and $\Delta\Omega_c(T)=-4\sigma^2/\epsilon (T-T_0)$. Since the size of the critical germ $L_c$ decreases as $T$ is lowered below $T_0$, our model assumes also a decrease with temperature of the maximum intrinsic size of the final plates (more precisely a proportionality is assumed $L=nL_c$, with $n$ to be determined from the condition that the first formed plate has the maximum size imposed in the simulation). Importantly, depending on the randomly chosen nucleation place, some plates may not be allowed to grow to the maximum size for a given temperature and only develop to a smaller size. This mechanism is responsible for the memory properties of the model.
Next, the heat flow is modeled towards a reservoir whose temperature linearly decreases, as in typical DSC experiments, the parameters being chosen so that the heat is not entirely used for the pahse transition (latent heat) but also the sample gradually cools during the austenite-martensite transformation.

\begin{figure}[ht]
\centering
\hskip -0.3cm
\includegraphics[scale=0.5]{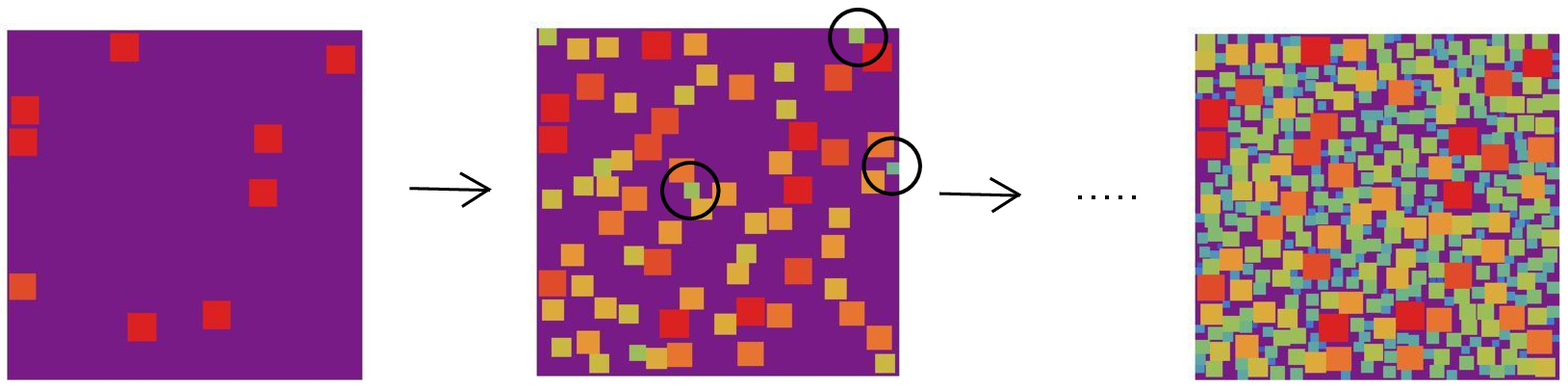}
\includegraphics[scale=0.37]{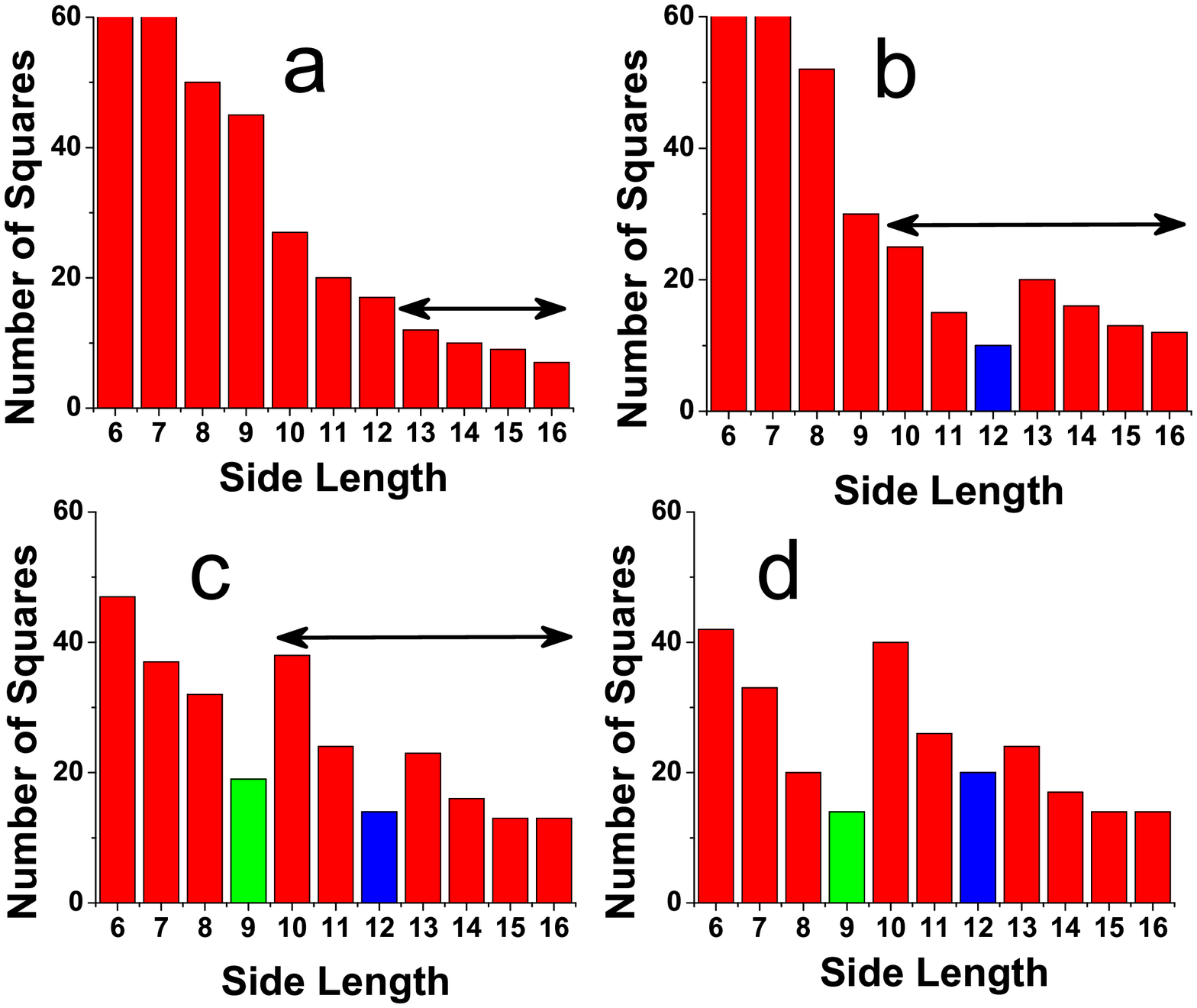}
\vskip -6cm
\caption{
{\it Upper part}: Martensitic plates nucleating randomly and growing to a maximum size, which decreases with temperature (from left to right). In the middle panel, the three black circles indicate plates that grow to lower sizes due to geometrical constrictions, which is a key mechanism (see description in text). {\it Lower part}:
(a) Plates sizes distribution after a "normal" martensitic transformation, starting from {\it zero} pre-existing plates -only squares of sides bigger than $6$ are given, for clarity-. Now let us assume an incomplete reverse transformation, which transforms back the plates of sizes $12$ and smaller. The double-head arrow indicates the untransformed plates. (b) Plates sizes distribution after a direct transformation starting with the untransformed plates at point a). Note the increased number of big size plates, but a depletion of plates of size $12$ (the blue bar), due to geometrical constrictions. (c) Plates sizes distribution after a direct transformation starting with the untransformed plates at point (b), of sizes higher than $9$. A depletion of sizes $9$ is noticed (the green bar), while the depletion of sizes $12$ is reduced. (d) By repeating the procedure from c), the {\it green} arrest point is accentuated while the number of {\it blue} squares increases further, thus "fading" the memory of the higher temperature.}
\end{figure}


  In order to capture multiple arrest points, a rather larger size of the system was simulated (a $200 X 200$ mesh) and also a big number of plates sizes were considered, ranging from square side $16$ -the largest- to side $3$ -the smallest-. The plates sizes distribution in Fig.3a is the "normal" distribution after a complete martensitic transformation. The parameters of the numerical simulation have been chosen to generate a balanced distribution, in the sense that the areas of the martensite phase being in the form of different sizes squares are roughly comparable (for instance there are 7 plates of side 16 with a total area of 1792, and 70 plates of side 6 with a total area of 2520, etc.). Next, an incomplete reverse transformation is assumed, during which the plates of side greater than 12 remain untransformed, as indicated by the arrow. The next direct transformation will start from this pre-existing puzzle of untransformed plates, and the effect will be a depletion of sample in square of side 12, marked with blue in Fig.3b, as they are most affected by geometrical constraints (a reduced effect being also on squares of side 11).

In the following, another incomplete reverse transformation is assumed, which this time leaves untransformed all plates of side greater than 9. A depletion of the sample in size 9 plates is noticed in Fig.3c (with color green), but also more "blue" squares have formed and the depletion of the sample such plates is reduced. Fig.3d shows that repeating the procedure accentuates the effect. Since, within the model, the thermal memory is associated with a depletion of certain martensite plates sizes, a reduction of the sizes depletion can be regarded  as a memory fading.

Finally, we simulate a reverse transformation, starting from the various plates sizes distribution presented in Fig.3a)-d). The DSC heat flow is modeled using the same parameters as for the direct transformation in some instances (as the specific heat, or the heat exchange rate with the reservoir), while the nucleation rate is replaced by the rate with which the martensite plates transform back $J_{M\rightarrow A}= J'_0\theta(T-T_L)f(L,T)$, with the notations from \cite{Tolea1}, the essential quantity being the Heaviside step function which assures that the transformation back for each size of plates can start only after they become thermodynamically unstable (at the temperature $T_L$), and the stress accumulated at border overcomes the austenite-martensite volume enthalpy difference: $T_L=T_0-4\sigma'/(\epsilon L)$. Necessarily, one must have $\sigma '\leq \sigma$, where the previously introduced $\sigma$ is the energy penalty at the border between phases and is usually only partially of elastic (non-disipative) nature. The results of the simulations are presented in Fig.4, with the panel a) corresponding to the transformation back on a previous complete direct transformation, and starting from the plates distribution from Fig.3a. The small oscillations noticed do not have a particular physical meaning, rather they are a result of the finite number of sizes considered. Each small dip corresponds to another plates sizes beginning the transformation back. As such, if certain sizes are depleted, as in Fig.3b (after a previous incomplete reverse transformation) the DSC signal shows a specific dip, marked by the blue circle in Fig.4b. If two arrests are performed in decreasing order of temperatures, resulting the plates distribution from Fig.4c, our simulation of the reverse transformation shows two slight dips (not very clear) marked by the blue and green circles respectively -Fig.3c. If the arrest at the "green" temperature is repeated, and we simulate the transformation back of the plates distribution in Fig.4c, a clear dip is observed in Fig.4d, indicated by the green circle, while the "blue" dip is completely vanished.

Therefore the simulation results presented in this Section reproduce quite well the memory fading experimentally found in the previous Section.


Since our model basically consists in randomly filling a surface with squares of different sizes, it is interesting to mention that there will also remain some untransformed area (where not even the smallest plates can fit), which is as if Martensite final temperature is not reached. By increasing the number of plates sizes, for the situation in Fig.3 being considered 14 different sizes, the untransformed surface reduces to few percents, closer to the experimental situation.
The problem of average filling of a surface (or volume) with randomly placed objects is common to the applied statistical geometry (see, e.g. \cite{YZheng,Ammi,HFJ,FW}), and while it is not a main focus of the present paper, it shall be addressed in a future work, for our 2D model.

 \begin{figure}[ht]
\centering
\vskip 0.1cm
\hskip 0.1cm
\vskip -0.5cm
\includegraphics[scale=1]{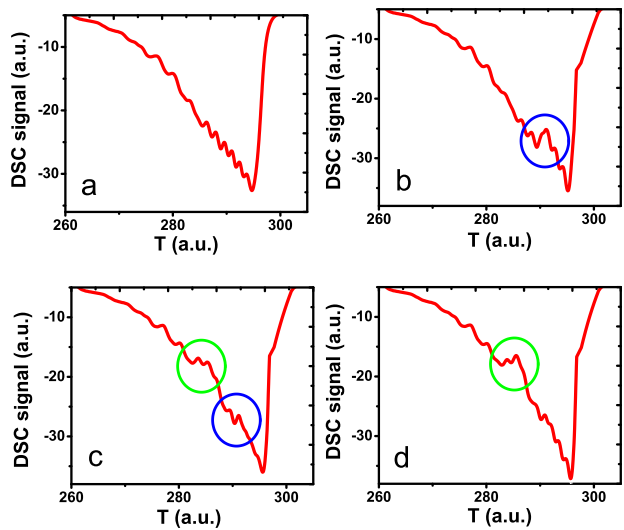}
\vskip -18.5cm
\caption{Simulation of the calorimetric signal for the reverse transformation. Panels a), b), c) and d) respectively correspond to the starting plates sizes distributions from the Panels with the same letter from Fig.3. In (a), the DSC signal following previous complete phase transitions shows no noticeable large dip. Next, note that the "blue" dip corresponding to an initial -higher- memorized temperature in (b) gets much reduced after recording the "green" arrest point in (c) and completely vanishes after recording the "green" arrest twice (d) - while the "green" dip itself accentuates.}
\end{figure}


\section{Conclusions and further discussions}


Two arguments are presented in support of a thermal memory fading effect in shape memory alloys, which takes place by heating to an {\it lower} temperature than the initially memorized one (previous studies addressed the memory erasing by heating to a {\it higher} temperature \cite{Airoldi1,WANG_SM,R-AACTA2,Wang-IJSNM,Zhou}):

\begin{itemize}
\item[(a)] The first argument is experimental, showing the effect in polycrystalline NiFeGa ribbons (we mention that in the bulk counterpart no thermal memory properties were found). If a first incomplete martensite-austenite transformation is stopped (arrested) at $T_1$ and a second one at a nearby temperature $T_2<T_1$, then the next full reverse transformation will show a reduction in amplitude of the calorimetric dip corresponding to $T_1$, together with the expected new dip corresponding to $T_2$. Repeating the arrest at $T_2$ leads to further reduction, or even disappearance, of the dip corresponding to $T_1$ (while the dip at $T_2$ accentuates). The effect is stronger if the difference $T_1-T_2$ is smaller.
 \item[(b)] The second argument is a 2D numerical simulation, previously proposed in \cite{Tolea1}, which associates the thermal memory with the depletion of the martensite plates of certain size(s), this depletion being an effect of geometrical constraints. First, it is shown that the model reproduces the memorizing of two temperatures, in decreasing order $T_2<T_1$. Secondly, it is shown that the arrest at $T_2$, performed once or several times (for a stronger effect) leads to a reduction of the depletion of sizes corresponding to $T_1$, thus a "fading" of the memory associated with this temperature. Simulated DSC curves reproduce qualitatively the experimental ones.
\end{itemize}

The theoretical modeling in (b) is phenomenological, so it can not be regarded at this point as a direct (or the only possible) explanation for the experimental findings from (a). Nevertheless, since the model reproduces well the previously established phenomenology of the thermal memory effect (see points (i),(ii) and (iii) from introduction) it brings plausibility arguments also for the novel memory fading effect.


A comment is in order: some previous experiments addressing multiple temperature arrest points, recorded in decreasing order (e.g.  \cite{WANG_SM,Wang-IJSNM,Zhou}), could in principle have exhibited the discussed memory fading effect, but this aspect was not specifically emphasized, to our knowledge. The lack of such effect in \cite{WANG_SM}, for instance, may be due to the large difference between the arrest points of about $3^0C$, while we find noticeable memory fading effects for arrest points closer than $2^oC$. In \cite{Wang-IJSNM} (addressing TiNiCu thin films) however, the data from Fig.7 with consecutive arrest temperatures at $1.5^oC$ or less apart may be interpreted as a memory fading effect of the higher temperatures (i.e. the calorimetric dip reduces in amplitude) due to subsequent recording of the lower ones. Some data from \cite{Zhou} (see, e.g. Fig.2), on NiMnGa based alloys may also be interpreted in this sense.

A potentially important application of the TME effect in SMA was proposed in \cite{Tang}, consisting in NiTi based thermometers for recording the maximum reached temperature during on overheating process, if it is in the range of the phase transition. For such an application, the memory fading effect discussed in this paper may introduce certain limitations, in the sense that the memory of the maximum reached temperature may be erased by (repeated) heating to a (nearby) lower temperature.


\section{Acknowledgements}

We acknowledge support from the Romanian Ministry of National Education,  grants PN-II-ID-PCE-2012-4-0516 and
Core Program PN16-480103.




\vskip 1cm
*corresponding author e-mail: tzolea@infim.ro

\end{document}